\documentclass[twocolumn,showpacs,preprintnumbers,amsmath,amssymb,prl]{revtex4}

\usepackage{graphicx}
\usepackage{bm}

\begin{document}

\title{Thermodynamics and Widom lines in supercritical carbon dioxide}

\author{Yu. D. Fomin}
\affiliation{Institute for High Pressure Physics, Russian Academy
of Sciences, Troitsk 142190, Moscow, Russia \\ Moscow Institute of
Physics and Technology, Dolgoprudny, Moscow Region 141700, Russia}
\author{V. N. Ryzhov}
\affiliation{Institute for High Pressure Physics, Russian Academy
of Sciences, Troitsk 142190, Moscow, Russia \\ Moscow Institute of
Physics and Technology, Dolgoprudny, Moscow Region 141700, Russia}
\author{E. N. Tsiok}
\affiliation{Institute for High Pressure Physics, Russian Academy
of Sciences, Troitsk 142190, Moscow, Russia}
\author{V. V. Brazhkin}
\affiliation{Institute for High Pressure Physics, Russian Academy
of Sciences, Troitsk 142190, Moscow, Russia}
\author{K. Trachenko}
\affiliation{School of Physics and Astronomy Queen Mary University
of London, Mile End Road, London E1 4NS, United Kingdom}

\date{\today}

\begin{abstract}
Behavior of supercritical fluids attracts a lot of attention
nowadays. It is important both from the point of view of
fundamental science and technological applications. However, up to
now the progress in the field is rather moderate. In this article
we report a computational study of supercritical carbon dioxide
which is one of the most important fluids for chemical industry.
We study the response functions of $CO_2$ in supercritical regime
and calculate the locations of their maxima (Widom lines). We also
make preliminary calculations of the line of crossover of
microscopic dynamics of particles (Frenkel line). The conclusions
on the Frenkel line location can be applied to study of the
atmosphere of Venus.
\end{abstract}

\pacs{61.20.Gy, 61.20.Ne, 64.60.Kw} \maketitle

\section{I. Introduction}

It is well known that many thermodynamic functions demonstrate
maxima in the vicinity of a critical point. Among these values
are, for example, heat capacity, isothermal compression, heat
expansion coefficients etc. Investigation of supercritical maxima
gave an incredible effect on a phase transition theory and in
particular the theory of critical phenomena. However, it also has
an important technological impact. Supercritical liquids can be
used in chemical industry as they have high solubility, rate of
chemical reactions etc.

Recently it was proposed that the locations of maxima of different
quantities lie not far from each other in $\rho - T$ or $P -T$
plane. In this case one introduces a single line of supercritical
maxima. This line was named as a Widom line \cite{widomstanley}.
However, later it was shown that even in the simplest liquids like
van der Waals \cite{widomvdw}, Lennard-Jones (LJ)
\cite{widomlj,may}, square well (SW) \cite{widomsw} and a toy
model studied in the work \cite{widomindusy} maxima of different
thermodynamic functions can be rather far from each other.
Moreover, the lines of maxima of different quantities behave
qualitatively different: some lines go to higher densities as
temperature increases while others go to lower densities with
increasing temperature. As a result all the curves rapidly diverge
and form a wide bunch rather than a line. In Refs.
\cite{imre,franzesestanley,vega} Widom line of water was
considered and it was shown that the qualitative behavior of
supercritical maxima of water looks very similar to the one of the
model liquids mentioned above. However, the locations of the
maxima points in $\rho - T$ or $P-T$ planes are reported in
literature only for a few systems
\cite{widomvdw,widomlj,may,widomsw,widomindusy,imre}.



Some system can demonstrate multiple Widom lines. One can expect
the appearance of multiple Widom lines in the systems with complex
phase diagrams like water. In particular, multiple Widom line
looks natural in the systems with liquid-liquid phase transition.
According to recent studies carbon dioxide can also possess very
complex phase diagram which includes molecular, polymeric and
dissociated metallic fluid phases \cite{bonev}. In addition to
conventional liquid - gas Widom line of $CO_2$ in this paper the
authors report the Widom line appearing due to the transition
between molecular and polymeric liquids calculated by isothermal
compressibility maxima. One can expect that the qualitative
behavior of this second Widom line of carbon dioxide will be
similar to the behavior of the usual gas - liquid Widom line.
However, this question requires further clarification.

Another line which divides a fluid in two domains -
low-temperature "rigid" one and high-temperature non-rigid fluid
was proposed in the literature \cite{ufn,frpre,frprl,bolm}. It was
named as Frenkel line. Frenkel line separates fluid by means of
its microscopic dynamics. In case of rigid regime the particles of
liquid perform several oscillations around an equilibrium position
and then the equilibrium position changes. In non-rigid regime the
oscillations are not observed.

Several methods to locate the Frenkel line in $\rho - T$ or $P -
T$ plane were proposed. Among them two the most convenient are
based on velocity autocorrelation function and isochoric heat
capacity.

Oscillations of the particle motion in rigid fluid are easily
recognized in velocity autocorrelation function by its
non-monotonous behavior. In case of non-rigid regime the velocity
autocorrelation function simply monotonically decay to zero.

The isochoric heat capacity criterium is based on the following
reasoning. Consider a monatomic system. The heat capacity of
monatomic crystal (in units of $k_B$) is equal to $3$. The heat
capacity of crystal is defined by both longitudinal and transverse
excitations. The difference between rigid and non-rigid fluids is
that the former can sustain transverse excitations while the later
cannot. The contribution of transverse excitations into the heat
capacity is $1$, so the crossover line is defined as $c_V=2$.

This article presents a computational study of supercritical
carbon dioxide $CO_2$. We report the thermodynamical properties of
$CO_2$ (equation of state and response functions) and location of
supercritical maxima in $\rho - T$ plane. This particular system
was chosen for its importance in chemical industry and planetary
science. We investigate the behavior of the heat capacity, the
isobaric expansion, the isothermal compressibility and the density
fluctuations in the framework of computer simulation. We also
perform a preliminary study of Frenkel line of carbon dioxide.

\section{II. System and Methods}

We studied the supercritical maxima of $CO_2$ by means of
molecular dynamics simulation. The model potential proposed in
Ref. \cite{pot} was used. This potential was optimized for
simulation of the liquid-gas coexistence, and it gives the
saturation curve and critical point in close agreement with
experimental data.

In this model carbon dioxide molecules are considered as rigid
bodies, i.e. $C-O$ bond length and $O-C-O$ angle are fixed. It has
an important consequence for heat capacity. One can show that the
heat capacity of ideal gas of rigid triatomic molecules is $4.5$
(here and later talking about heat capacity we implicity assume
that energy and temperature are measured in the same units
($k_B=1$), so heat capacities are dimensionless). The value $3.0$
comes from thermal motion of the atoms and $1.5$ from rotation of
the molecules.


A system of $4000$ $CO_2$ molecules was simulated in a cubic box
with periodic boundary conditions in canonical ensemble (fixed
number of particles $N$, volume of the system $V$ and temperature
$T$). The timestep was set to $0.1$ fs. A period of $100$ ps was
used to equilibrate the system. Then the system was simulated for
$100$ more ps in order to compute the averages. So short
simulations were enough due to high temperature regimes studied in
the work. The system was simulated at the densities from
$0.2g/cm^3$ up to $0.7 g/cm^3$ with the step $\Delta \rho=0.025
g/cm^3$. The temperatures studied were from $310K$ up to $400K$
with step $dT=10K$ and then up to $900K$ with step $20K$. Critical
parameters of the model were reported in Ref. \cite{pot}:
$T_c=304K$, $\rho_c=0.467 g/cm^3$ and $P_c=7.23 MPa$. So the
parameters used in our work correspond to the temperatures from a
bit above $T_c$ to approximately $3T_c$ and from the densities
from approximately $0.5 \rho_c$ up to almost $3.5 \rho_c$. In
total $29$ different densities and $35$ different temperatures
were studied. It gives us large data set of energies and pressures
as functions of $\rho$ and $T$. These data were fitted to
polynomial functions of the form $a_{p,q} \rho^p T^q$. The fitting
parameters for pressure and energy are given in the table I. These
fitting functions were used to compute the thermodynamic response
functions: density fluctuations $ \left ( \frac{\partial
\rho}{\partial P} \right ) _T$, isothermal compressibility
$\chi_T= \frac{1}{\rho} \left ( \frac{\partial \rho}{\partial P}
\right ) _T$, thermal expansion coefficient
$\alpha_P=\frac{1}{\rho} \left ( \frac{\partial \rho}{\partial T}
\right ) _P$, constant volume heat capacity $c_V= \left (
\frac{\partial E}{\partial T} \right ) _V$ and constant pressure
heat capacity $c_P= \left ( \frac{\partial H}{\partial T} \right )
_P$ where $E$ and $H=E+PV$ are internal energy and enthalpy of the
system.


We also calculate the Frenkel line of the system. For doing this
we employ isochoric heat capacity method described in
Introduction. However, since carbon dioxide is not monotonic the
threshold values have to be modified. In case of rigid threeatomic
molecules the heat capacity criterium gives that the Frenkel line
corresponds to $c_V=3.5$. Here we neglect the anharmonic effects.
Although this is a strong assumptions we believe that it does not
affect the qualitative behavior of the system. This criteria is
used in the present article for preliminary estimation of the
Frenkel line location. In order to compute the Frenkel line the
simulations were extended up to the densities $\rho =1.5 g/cm^3$.

All simulations were performed using lammps simulation package
\cite{lammps}.

\begin{table}

\begin{tabular}{|c|c|c|c|}
  \hline
  p & q & $a_{p,q}$ for P & $a_{p,q}$ for En \\
  \hline
  0 & 0 & -59.199 & 8884.811 \\
  1 & 0 & -12.402 & -5347.037 \\
  2 & 0 & -1889.319 & 4859.453 \\
  3 & 0 & 4090.029 & 823.207 \\
  4 & 0 & -7561.170 & -1968.605 \\
  5 & 0 & 3396.4327 & 286.739 \\
  0 & 1 & 0.365 & 16.137 \\
  1 & 1 & 1.041 & 18.175 \\
  2 & 1 & -3.832 & -23.120 \\
  3 & 1 & 4.0548 & 2.038 \\
  4 & 1 & 4.704 & 2.006 \\
  0 & 2 & $-4.726 \cdot 10^{-4}$ & $-4.598 \cdot 10^{-2}$ \\
  1 & 2 & $7.126 \cdot 10^{-3} $ & $-2.406 \cdot 10^{-2}$ \\
  2 & 2 & $7.367 \cdot 10^{-3}$ & $3.183 \cdot 10^{-2}$ \\
  3 & 2 & $-5.046 \cdot 10^{-3}$ & $-3.640 \cdot 10^{-3}$ \\
  0 & 3 & $-1.354 \cdot 10^{-6}$ & $7.573 \cdot 10^{-5}$ \\
  1 & 3 & $-1.093 \cdot 10^{-5}$ & $9.430 \cdot 10^{-6}$ \\
  2 & 3 & $-9.953 \cdot 10^{-7}$ & $-1.315 \cdot 10^{-5}$ \\
  0 & 4 & $3.128 \cdot 10^{-9}$ & $5.926 \cdot 10^{-8}$ \\
  1 & 4 & $4.594 \cdot 10^{-9}$ & $9.248 \cdot 10^{-10}$ \\
  0 & 5 & $-1.614 \cdot 10^{-12}$ & $1.779 \cdot 10^{-11}$ \\
  \hline
\end{tabular}

\caption{Fitting coefficients for pressure and internal energy.
The fitting formula is $X=a_{p,q} \rho^p T^q$, where $X=P,En$. The
densities ranging from $0.2$ up to $0.7$ are used for this fit.}
\end{table}

\section{Results and Discussion}

\begin{figure}
\includegraphics[width=7cm, height=7cm]{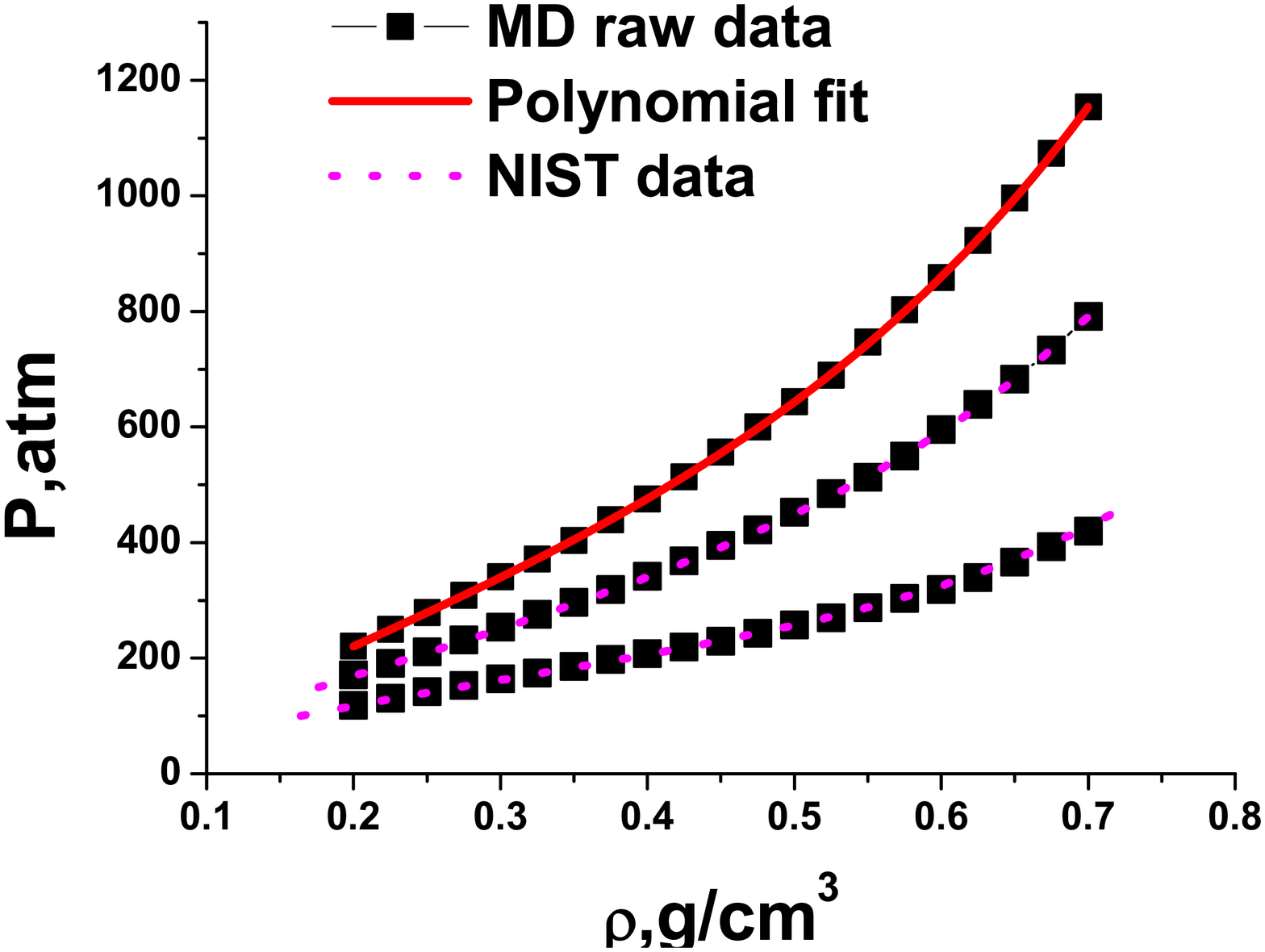}%

\caption{\label{fig:fig1} Equation of state of $CO2$. The plot
shows raw data from MD simulation (symbols), polynomial
approximation for $T=600K$ and experimental data from NIST
database for $T=400K$ and $500K$. Temperature from bottom to top:
$400K$, $500K$ and $600K$(Color online).}
\end{figure}

First we report the equations of state of the supercritical carbon
dioxide (Fig.~\ref{fig:fig1}). This figure shows our MD data in
comparison to experimental results from NIST database \cite{nist}.
One can see that the MD data are in good agreement with
experimental results. The figure also demonstrates the polynomial
fit of the data. These fit also well describe the molecular
dynamics results

An important conclusion from Fig.~\ref{fig:fig1} is that the
results of simulations are in good agreement with experimental
results which justifies the reliability of our study. It means
that even if the potential we employ in the present study
\cite{pot} was fitted to reproduce the liquid-gas saturation line
it can be also successfully used to study the system and analyze
its properties in supercritical region.

\begin{figure}

\includegraphics[width=7cm, height=7cm]{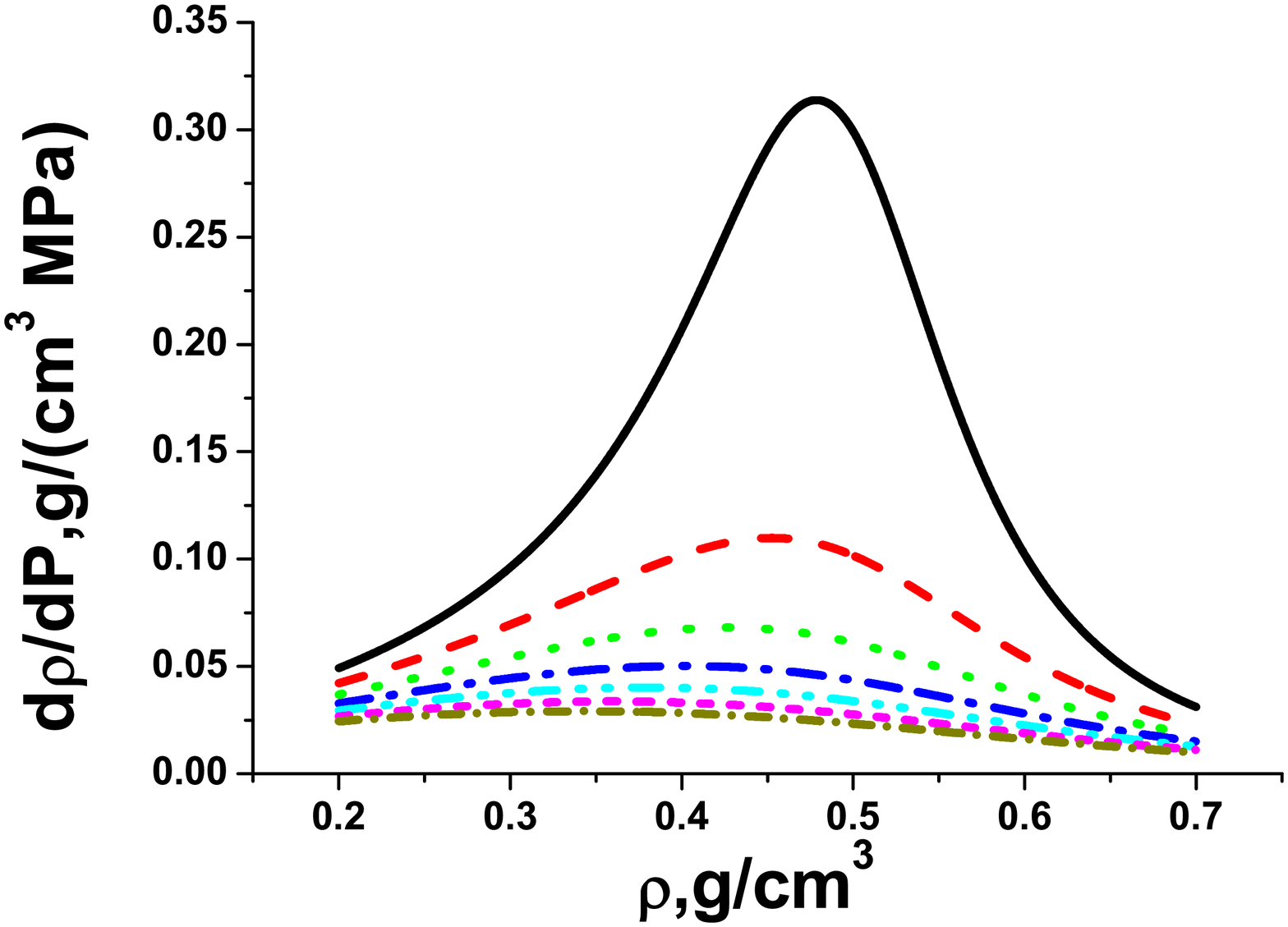}%

\caption{\label{fig:fig2} Density fluctuations along several
isotherms. The temperatures are from $T=320K$ (top) up to $T=380K$
(bottom) with step $\Delta T=10K$. (Color online).}
\end{figure}

Fig.~\ref{fig:fig2} demonstrates the density fluctuations at a set
of temperatures slightly above the critical one. One can see that
at the lowest temperatures there is a well defined maximum.
However, this maximum rapidly decays. In fact, the maximum is
observed even at quite high temperatures (as high as $540K$), but
it becomes of the order of numerical errors.

\begin{figure}
\includegraphics[width=7cm, height=7cm]{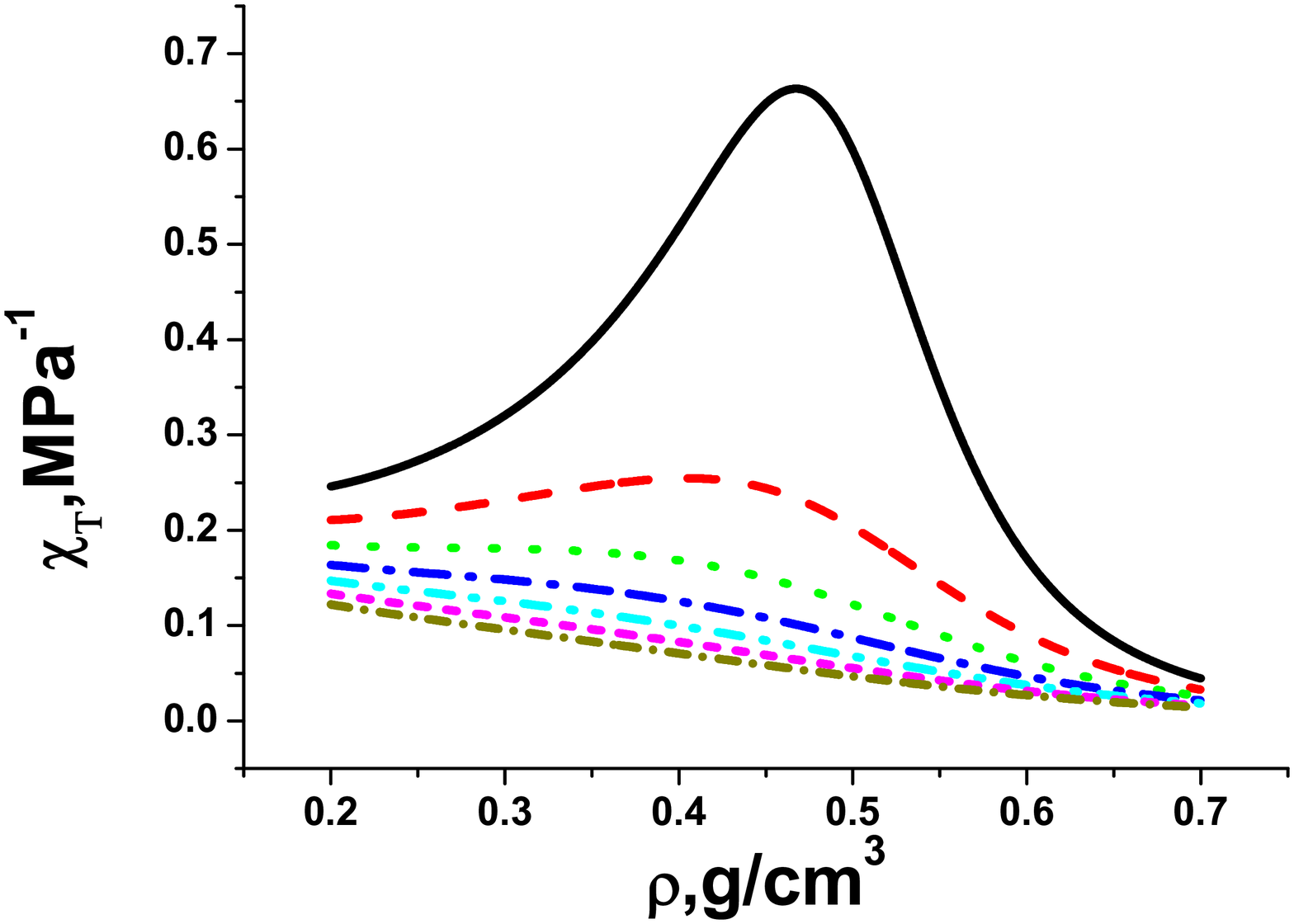}%

\caption{\label{fig:fig3} Isothermal compressibility of $CO_2$
along several isotherm. The notation is the same as for
Fig.~\ref{fig:fig2}. (Color online).}
\end{figure}

Fig.~\ref{fig:fig3} shows the behavior of isothermal
compressibility at the same set of temperatures. Unlike the
previous case the maximum quickly vanish. Even at the temperature
as low as $350K$ it is already unobservable.

\begin{figure}
\includegraphics[width=7cm, height=7cm]{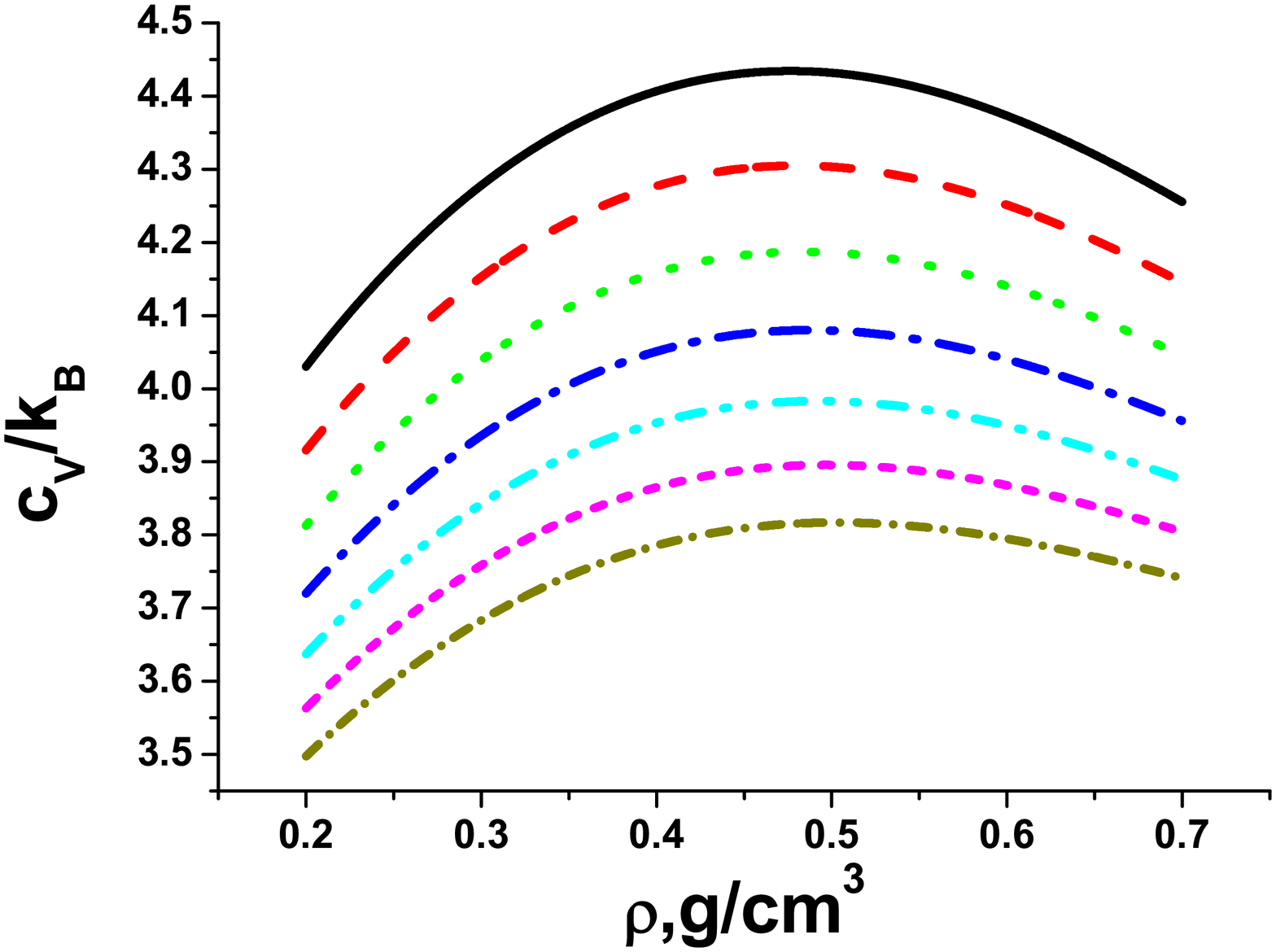}%

\caption{\label{fig:fig4} Isochoric heat capacity $CO_2$ along
several isotherm. The notation is the same as for
Fig.~\ref{fig:fig2}. (Color online).}
\end{figure}

Fig.~\ref{fig:fig4} demonstrates the isochoric heat capacity along
the same set of isotherms. Again there are well defined maxima.
The maxima can be traced to the temperatures as high as $440K$.

\begin{figure}
\includegraphics[width=7cm, height=7cm]{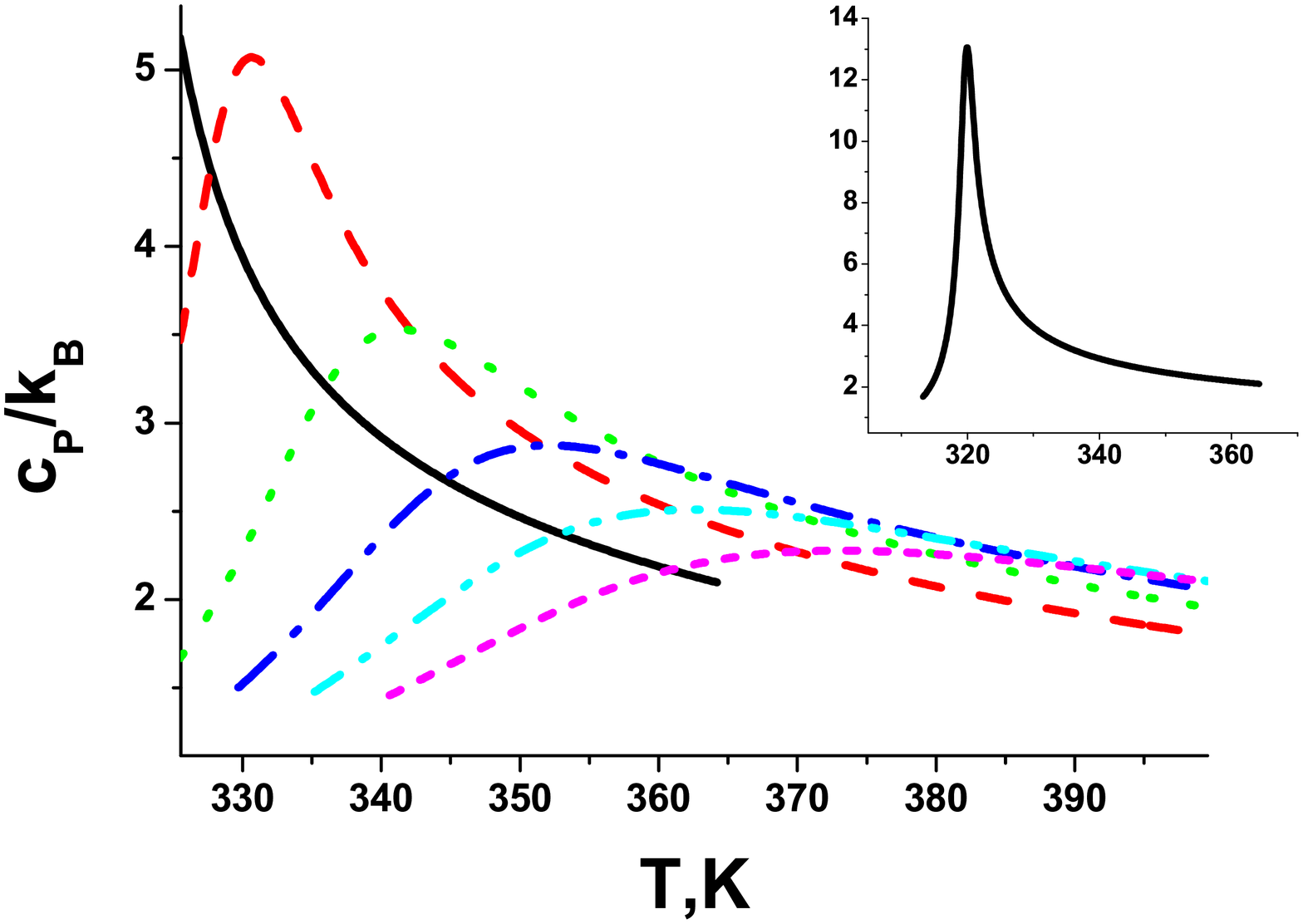}%

\caption{\label{fig:fig5} Isobaric heat capacity along a set of
isobars. Pressures are from $10MPa$ (left) up to $20MPa$ (right)
with step $\Delta P=2MPa$. The inset enlarges the plot for
$10MPa$. (Color online).}
\end{figure}

The behavior of isobaric heat capacity $c_p$ was studied both
along isobars and isotherms. Fig.~\ref{fig:fig5} shows that $c_p$
demonstrates a sharp maxima when approaching $T_c$ along the
isobars. As pressure increases the maximum becomes broader and
less pronounced. It can be observed even at the pressures up to
$100MPa$. However, at so high pressures the maximum becomes of the
order of the numerical errors.

\begin{figure}
\includegraphics[width=7cm, height=7cm]{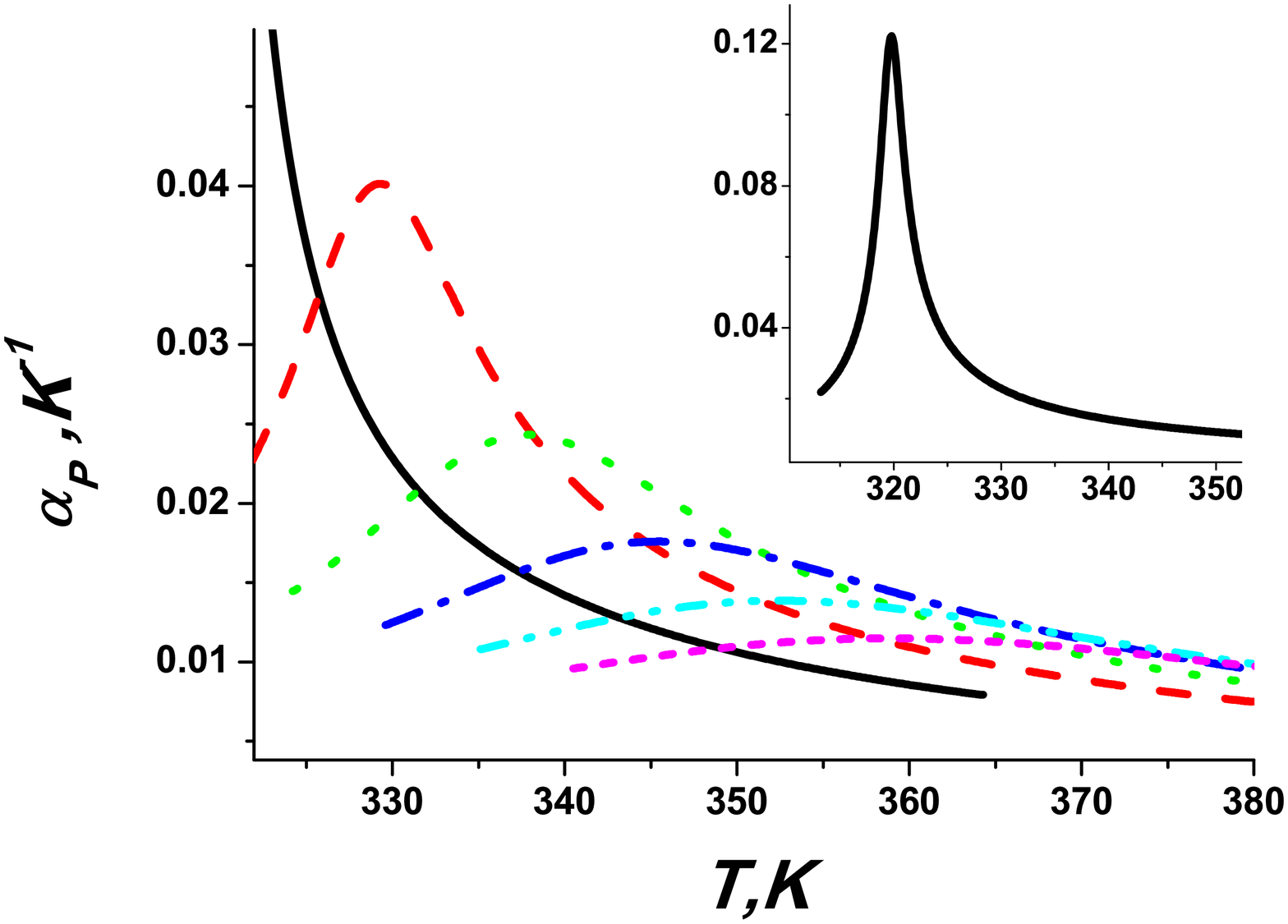}%

\caption{\label{fig:fig6} Isobaric expansion coefficient along a
set of isobars.Pressures are from $10Mpa$ (left) up to $20MPa$
(right) with step $\Delta P=2MPa$. The inset enlarges the plot for
$10MPa$. (Color online).}
\end{figure}

The last quantity studied is isobaric expansion coefficient
(Fig.~\ref{fig:fig6}). This coefficient was also calculated both
along isobars and isotherms. Similar to $c_p$ it also demonstrates
sharp maxima at temperatures close to $T_c$. These maxima rapidly
decrease with increasing pressure. The largest pressure we were
able to observe the maximum is $44.0 MPa$.

\begin{figure}
\includegraphics[width=7cm, height=7cm]{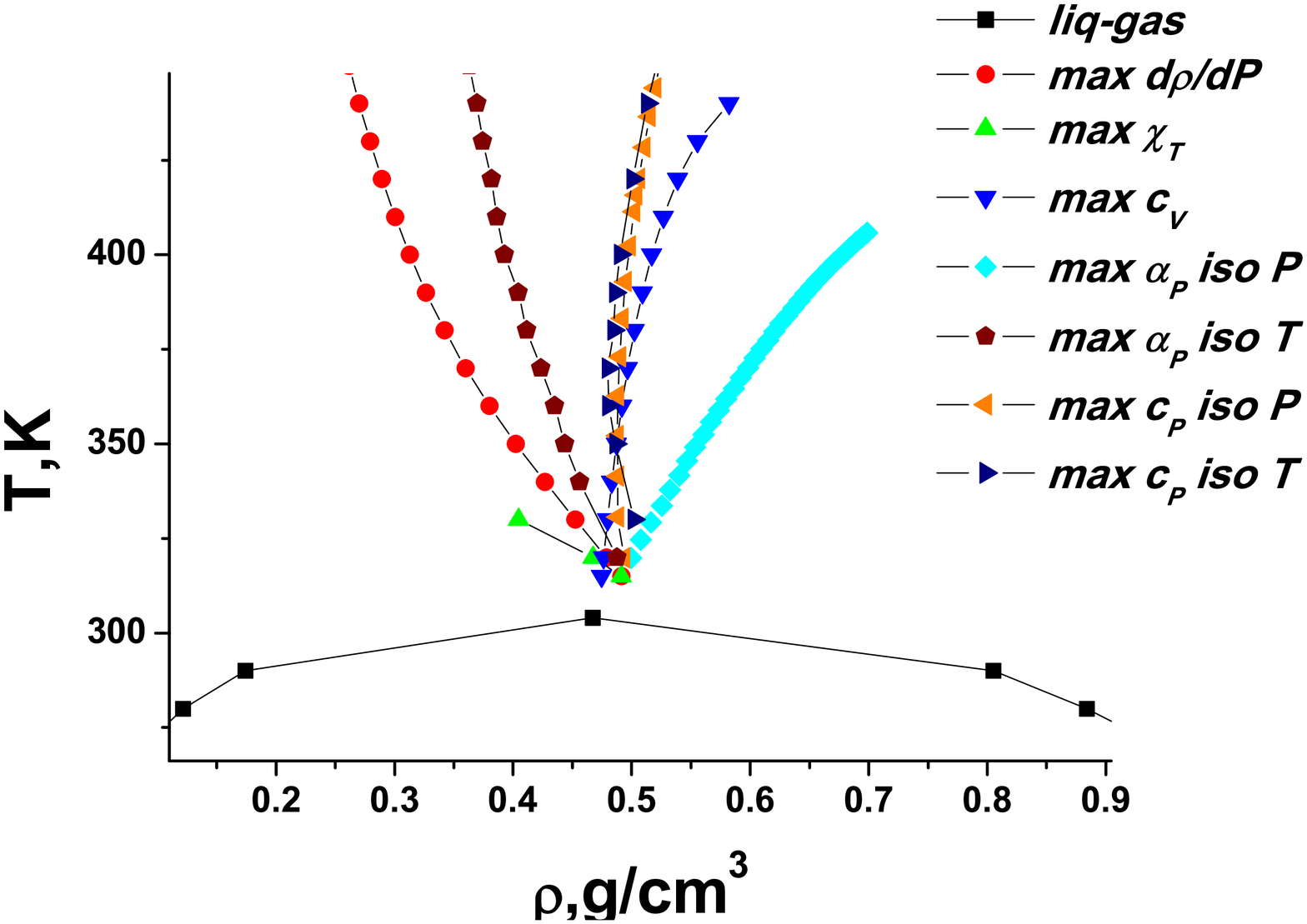}%

\caption{\label{fig:fig7} Location of maxima of different
thermodynamic functions of $CO_2$ in $\rho - T$ plane. The
liquid-gas coexistence curve is taken from Ref. \cite{pot}. (Color
online).}
\end{figure}

The location of maxima of different quantities is shown in
Fig.~\ref{fig:fig7}. One can see that the qualitative behavior of
different quantities is analogous to the case of LJ fluid. The
fluctuations of density and isothermal compressibility maxima move
to the lower densities with increasing temperature while heat
capacities and isobaric expansion coefficient maxima shift to
higher densities at higher temperatures. As a result the lines of
maxima rapidly diverge forming a bunch of lines. For example,
already at $T=320K$ ($4 \%$ above the critical point) the
difference between the densities of maxima of different quantities
is as large as approximately $12 \%$. So one can estimate that the
lines of maxima of different quantities fall onto the same line
only at very small temperatures above the critical one.

Importantly, even maxima of the same quantity taken along
different thermodynamic trajectories (isotherms, isochors,
isobars) can have different location. One can see, that the maxima
of $\alpha_P$ along isotherms and isobars are very different. At
the same time the locations of isothermal and isobaric maxima of
$c_p$ are almost identical.

\begin{figure}
\includegraphics[width=7cm, height=7cm]{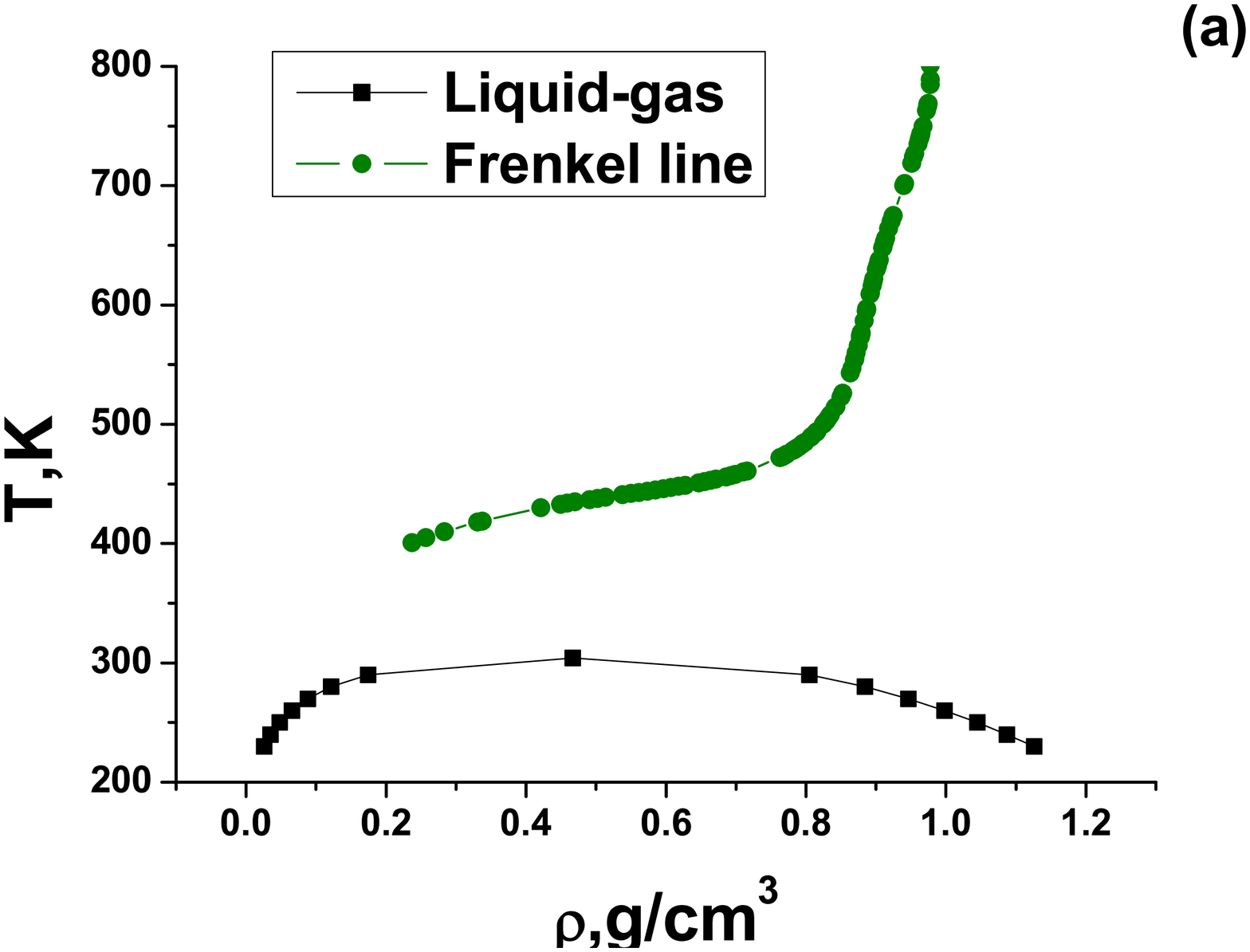}%

\includegraphics[width=7cm, height=7cm]{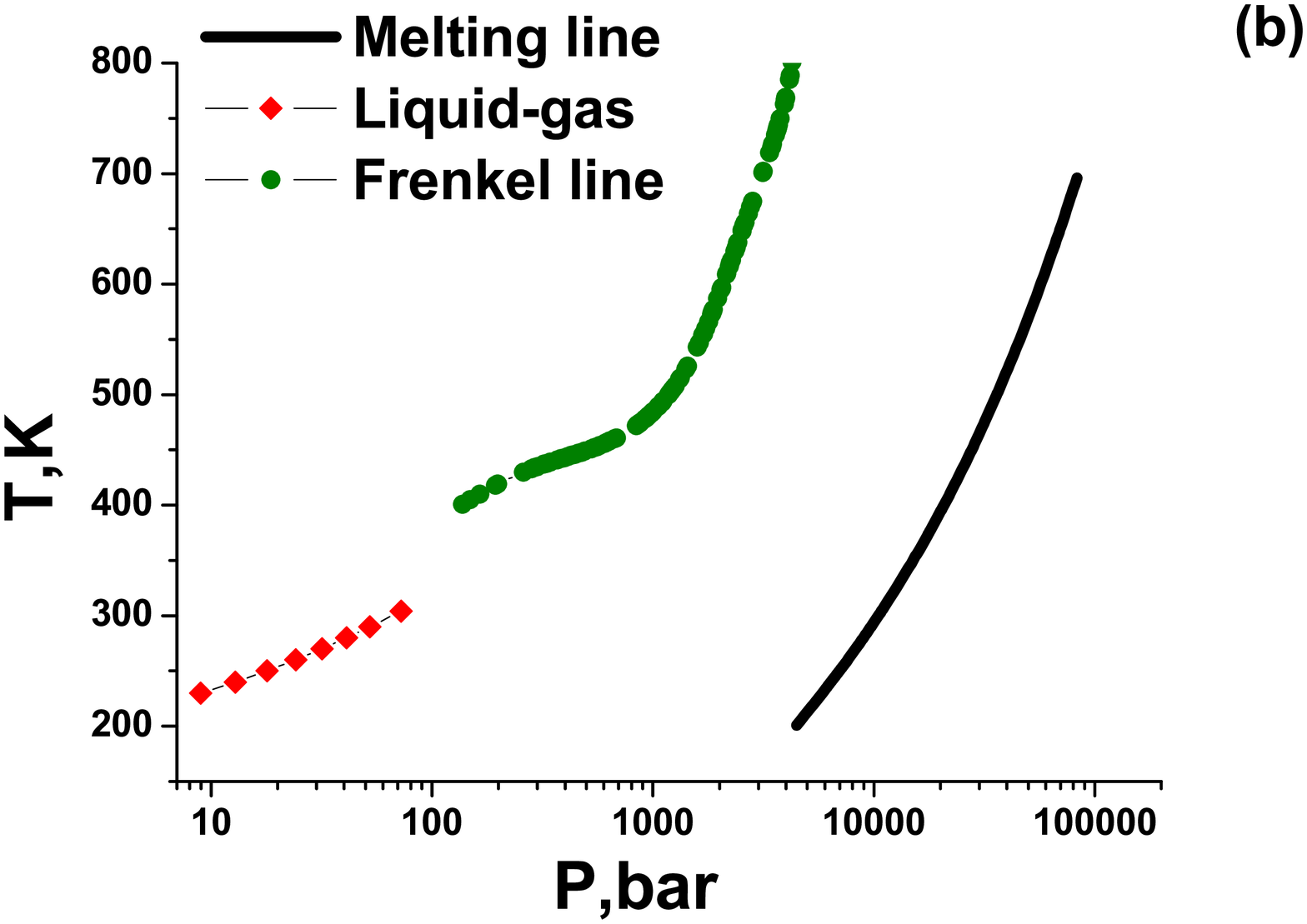}%

\caption{\label{fig:fig8} Location of Frenkel line of $CO_2$ in
(a) $\rho - T$ and $P-T$ planes. (Color online).}
\end{figure}

At last we present a preliminary estimation of Frenkel line of
$CO_2$. Several methods to find the location of Frenkel line were
proposed earlier \cite{ufn,frpre,frprl}. In the current
publication we use the method based on isochoric heat capacity
$c_v$. Figs.~\ref{fig:fig8} (a) and (b) show the location of
Frenkel line in $\rho - T$ and $P-T$ planes. In case of $P-T$
coordinates we add also an experimental melting line from Ref.
\cite{melting} (Eq. (3)).


One can see from the Figs.~\ref{fig:fig8} (a) and (b) that Frenkel
line calculated from $c_v=3.5$ criterium demonstrates a bend at
relatively low temperatures. This bend is related to supercritical
maximum of heat capacity and does not correspond to rigid -
non-rigid fluid crossover. At higher temperatures and pressures
Frenkel line becomes approximately parallel to the melting line in
logarithmic coordinates in $P$.

Investigation of $CO_2$ is important for planetary science and in
particular Venus. The concentration of carbon dioxide in the
atmosphere of Venus is approximately $96 \%$. The current
temperature of Venus is estimated as high as $735 K$ while
atmospheric pressure is $9.3 MPa$. One can see from Fig.
~\ref{fig:fig8} (b) that this (P,T) point lies deeply inside
gaseous phase. However, it is argued in the literature \cite{urey}
that at earlier stages the atmospheric pressure of Venus could be
as high as several dozen of MPa. It means that early Venus
atmosphere was consisted of rigid carbon dioxide fluid while later
on it experienced crossover into non-rigid state. This crossover
is related to change of microscopic dynamics of the fluid which
leads to altering many physical properties which could strongly
affect the formation of the Venus relief.

\section{VI. Conclusions}

We study the equation of state and superctiritical maxima of
carbon dioxide in molecular dynamics simulation. Comparing the MD
results with experimental data we conclude that they are in
excellent agreement. We show that similar to model liquids studied
before  the lines of supercritical maxima of $CO_2$ rapidly
diverge forming a wide bunch of lines. Moreover even the maxima of
the same quantity taken along different lines (isotherms or
isobars) form different lines. These results justify our earlier
conclusion that the concept of Widom line as the line of
supercritical maxima is ill defined and does not have clear
physical sense \cite{widomsw}.

In our recent publications we proposed that supercritical fluids
can exist in two states which differ in microscopic dynamics -
rigid and non-rigid fluid. These states are separated by so called
Frenkel line which can be determined from heat capacity criterium.
For rigid triatomic molecules this criterium approximately
corresponds to $c_v=3.5$ in units of $k_B$. We report the
calculations of Frenkel line of $CO_2$ and show that it goes
approximately parallel to the melting line in coordinates $logP -
T$.

\bigskip

\begin{acknowledgments}
Y. F. thanks the Russian Scientific Center at Kurchatov Institute
and Joint Supercomputing Center of Russian Academy of Science for
computational facilities. Y. F., E. Ts. and V. R. are grateful to
the Russian Science Foundation (Grant No 14-12-00820) for the
support. V. B. is grateful to the Russian Science Foundation
(Grant No 14-12-00093).
\end{acknowledgments}


\end{document}